\documentclass{PoS}
\newcommand{\hi}{{\sc H\,i} }

\title{Extragalactic VLBI surveys in the MeerKAT era}

\ShortTitle{Extragalactic VLBI surveys in the MeerKAT era}

\author{\speaker{Roger P. Deane}
        \\
       Rhodes University\\
       E-mail: \email{r.deane@ru.ac.za}}

\abstract{The past decade has seen significant advances in cm-wave VLBI extragalactic observations due to a wide range of technical successes,  including the increase in processed field-of-view and bandwidth. The future inclusion of MeerKAT into global VLBI networks would provide further enhancement, particularly the dramatic sensitivity boost to $>7000$~km baselines. This will not be without its limitations, however, considering incomplete MeerKAT band overlap with current VLBI arrays and the small (real-time) field-of-view afforded by the phased up MeerKAT array. We provide a brief overview of the significant contributions MeerKAT-VLBI could make, with an emphasis on the scientific output of several MeerKAT extragalactic Large Survey Projects.}

\FullConference{MeerKAT Science: On the Pathway to the SKA,\\
	25-27 May, 2016,\\
		Stellenbosch, South Africa}
		
\newcommand{\skipthis}[1]{}
\newcommand{\mnras}{MNRAS}

\newcommand{\aj}{AJ}
\newcommand{\apj}{ApJ}
\newcommand{\apjl}{ApJL}
\newcommand{\aap}{A\&A}

\newcommand{\pasp}{PSPS}

\newcommand{\nar}{New Ast. Reviews}

\newcommand{\pasa}{Publications of the Astronomical Society of Australia}

\begin{document}

\section{Introduction}\label{sec:intro}

The (sub)milli-arcsecond angular resolution afforded by a radio interferometer with continental baselines has yielded important results throughout the half century that the so-called Very Long Baseline Interferometry (VLBI) technique has been used (e.g., \cite{Thompson2001}). Scientific highlights include the measurement of apparent superluminal motion \cite{Whitney1971,Cohen1971}; measuring the expansion shells of supernovae (e.g., \cite{Bartel2002}); the discovery of a close binary supermassive black hole system \cite{Rodriguez2006}; time-variable accretion onto both micro-quasars (SS\,433, \cite{Vermeulen1993,Mioduszewski2004}) and supermassive black holes (M87, \cite{Walker2008}), the localisation of a Fast Radio Burst (FRB) \emph{within} its host galaxy \cite{Marcote2017}, as well as a direct measurement of the expansion rate of the Universe through modelling of disk water masers to extract a `standard ruler' (e.g. \cite{Reid2013}). Apart from the essential milli-arcsecond angular resolution that made these seminal observations possible, they share another common thread: each was made through serendipitous discovery or as part of dedicated VLBI followup, rather than a systematic wide-field VLBI survey. This is due to a variety of factors, such as limited sensitivity, Fourier coverage; processing power; software capabilities; and the small field-of-view limit imposed by time/bandwidth smearing. The latter was a practical necessity, particularly given the relative sparsity of the cm-VLBI sky at sensitivity limits of $\gg 1$~mJy\,beam$^{-1}$. 

As we look ahead to the SKA era, virtually all cm-wave radio telescopes are being designed to carry out wide-field surveys (e.g. MeerKAT, ASKAP, APERTIF, SKA1-MID, MFAA). It is natural to ask what will VLBI's role in this future be? Should it remain largely focused on the targeted and/or follow up observations that it has been so successful at performing; or should it attempt to move in step with the connected-element, arcsec-scale facilities in surveying large, contiguous areas of sky?  Further afield, if the long baseline ($\gg100$~km) component  of SKA Phase 2 is realised, then it will mark the point at which VLBI is merged into standard connected-element radio interferometry and carry out wide-field surveys by design (at milli-arcsecond angular resolution). Answers to the above questions are dependent on many aspects, including the required processing resources and the opportunity cost of forgoing current time on existing VLBI arrays. 

Despite the challenges of performing deep, high resolution radio observations over large areas, new software correlator developments \cite{Deller2011, Morgan2011} and larger instantaneous bandwidth have made this both more practical and more fruitful. This represents a dramatic shift in the efficiency of VLBI observations, and enables surveys of depth and area that will be a unique addition to multi-wavelength, wide-field surveys past, present and future. The technique is already being used to study multi-wavelength extragalactic fields at sensitivities down to the $\sim10~\mu$Jy\,beam$^{-1}$ level (e.g. \cite{Middelberg2011, Middelberg2013, Chi2013, Herrer2016}, Deane et al., in prep., Radcliffe et al., in prep.). However, the areas covered by these surveys are limited to $\Omega \lesssim 1$~deg$^2$, significantly less than the survey areas planned with next-generation instruments ($\gg 1$~deg$^2$). It will be highly challenging to perform sensitivity-matched VLBI surveys over comparative areas of sky unless (a) significantly more time is dedicated to adhoc VLBI arrays by participating observatories; (b) a number of new stations are added to the existing global networks; and/or (c) a significant fraction of the full field-of-view of new arrays such as MeerKAT can be processed (through offline beam forming, see Sec.~\ref{sec:offlinebeam}).

Given this context, these proceedings explore some of the scientific opportunities that MeerKAT offers towards this VLBI future, including wide-field mas-scale surveys of deep extragalactic fields. Such an objective would cement VLBI's role in `mainstream astrophysics' as urged by Malcolm Longair at the inaugural JIV-ERIC Symposium in April 2015. But perhaps more importantly, it would open up significant discovery space without having to rely on serendipity and/or pre-selection at another wavelength or with lower angular resolution radio observations. Scaling up to large-scale, wide-field VLBI on MeerKAT extragalactic fields opens three orders of magnitude in angular resolution discovery space; while processing $>4$ orders of magnitudes more of the available field-of-view of a typical VLBI observation. It also takes steps towards the post-processing paradigm that is required of the full SKA. As such, we consider both targeted and wide-field MeerKAT-VLBI surveys.

\section{VLBI-relevant attributes of MeerKAT}\label{sec:attri}

\begin{figure}[b]
\centering
\includegraphics[width=0.7\textwidth]{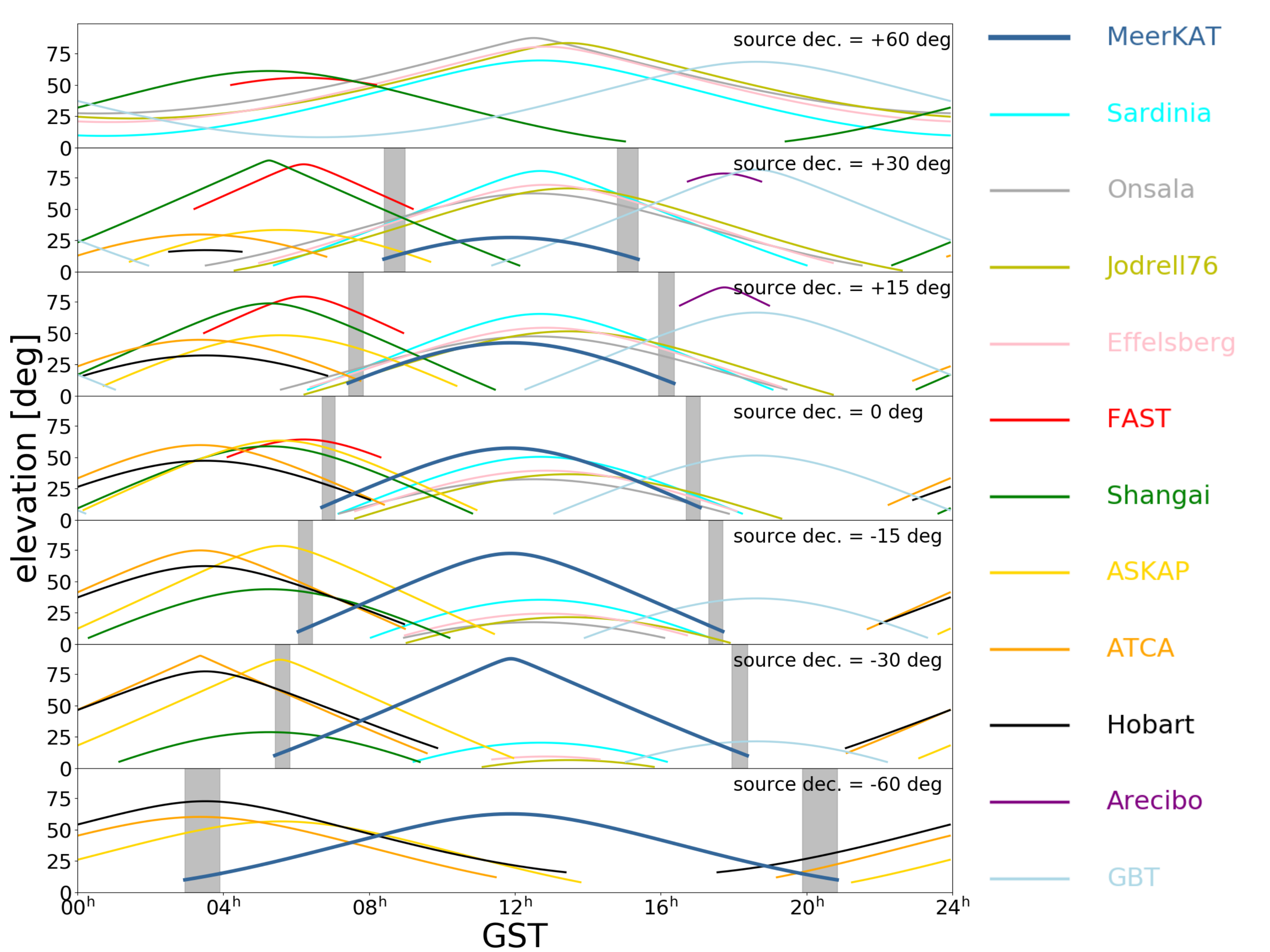}
\caption{Station elevation angles as a function of Greenwich Sidereal Time (GST), showing MeerKAT's (blue) mutual visibility with a selection of VLBI stations on the European, Asian, North American and Australian continents. The Right Ascension is arbitrarily chosen to place MeerKAT's maximum elevation at GST = 12$^{\rm h}$. The grey, vertical bands indicate the regions where MeerKAT's elevation is in the range of 10-15 degrees.  }
\label{fig:mutualvis}
\end{figure}

Here we consider MeerKAT's suitability with regard to participation in global VLBI networks. A phased up MeerKAT would have an effective area of a single $\sim110$~m antenna (with 1.4 GHz SEFD $\lesssim 10$~Jy) adding significant sensitivity to existing VLBI networks, but crucially at the longest baselines ($>7000$~km). For the European VLBI Network, MeerKAT would add this sensitivity in an important north-south direction and will strengthen the contribution of HartRAO through better sampling of and higher sensitivity in this part of the {\sl uv}-plane; as well as stronger constraints on absolute amplitude calibration. Similarly, for the Australian LBA it adds a sensitive 10,000 km east-west baseline. Importantly, it extends VLBI coverage in the southern hemisphere (see Fig.~\ref{fig:mutualvis}), which will be expanded even further by the African VLBI Network (AVN). 

MeerKAT band overlap with existing networks is relatively limited. In the L-band, MeerKAT's upper frequency limit of 1670~MHz (although the digitized band extends up to 1712~MHz), results in a common bandwidth of $\sim$80 MHz with typical EVN L-band observations (1590 \-- 1720 MHz). In S-band, there is 100~per cent frequency overlap ($\nu = 1750 - 3500$~MHz for MeerKAT), however, this is a significantly less sensitive EVN band (factor $\sim$3 difference). MeerKAT's X-band receivers (as yet unfunded) range from 8-14.5 GHz and thus would have full overlap with the standard EVN and Australian LBA X-band frequency setup. 

The phased up MeerKAT core (32 antennas within 1~km) results in a very narrow field-of-view ($\lesssim 0.5$~arcmin$^2$ at L-band). Since the number of simultaneous tied-array beams is 4, this would mean wide-field VLBI surveys would not benefit significantly from MeerKAT's high sensitivity and individual antenna's wide field-of-view, unless significantly more beams can be formed offline and independently correlated with other VLBI stations (more on this possibility in Sec.~\ref{sec:offlinebeam}). 

MeerKAT will be a heavily over-subscribed telescope. The array has $\sim$70\% of its time already allocated to the Large Survey Projects (LSPs) and will only operate for $\sim$5 years before merging with SKA1-MID. However, therein lies an opportunity given that MeerKAT will be able to simultaneously operate in interferometric and tied-array mode, with up to four simultaneous sub-arrays. There is a clearly an opportunity for commensal VLBI observations, which will undoubtedly enhance many of the MeerKAT LSPs science outputs as elaborated on in Sec.~\ref{sec:LSPs}.

\section{Extragalactic science opportunities with MeerKAT-VLBI}

\subsection{Targeted MeerKAT-VLBI observations}

The case for VLBI in the SKA era has been well captured in many previous works (e.g. \cite{Garrett2000, Fomalont2004, Godfrey2012, Paragi2015}). They outline the unique and important benefit VLBI provides not only radio astronomy, but the broader community as well. Naturally, much of the SKA1-MID VLBI discussion is directly relevant to MeerKAT-VLBI, which could be considered a SKA-VLBI precursor. The intention here is to highlight how MeerKAT can enhance approved extragalactic LSPs, however, there are of course a wide range of science cases that be significantly enhanced by MeerKAT-VLBI for the reasons outlined in Sec.~\ref{sec:attri}, specifically the long-baseline sensitivity, southern hemisphere coverage, and north-south (east-west) baselines. A few examples of science cases with MeerKAT-VLBI could make a \emph{critical contribution} include: the mas-scale investigation of southern or equatorial transient/variable sources (e.g. Fast Radio Bursts, \cite{Giroletti2016, Marcote2017}) detailed in MacLeod \& Bietenholz (these proceedings); high precision astrometry (e.g. \cite{Fomalont2004}); dark matter substructure in radio-loud, strong gravitational lenses \cite{McKean2015}; constraining the binary supermassive black hole inspiral rate (e.g. \cite{Burke-Spolaor2011, Deane2015}); and constraints on low frequency gravitational waves or anisotropic Hubble flow through increased proper motion coverage of southern hemisphere radio loud quasars (e.g. \cite{Gwinn1997, Titov2011}). As targeted VLBI observations are likely to continue to be the predominant mode of operation in the MeerKAT era, the above examples as well those outlined within \cite{Godfrey2012} and \cite{Paragi2015} stand to be significantly enhanced by MeerKAT participation in global VLBI networks.

\subsection{MeerKAT Large Survey Projects}\label{sec:LSPs}

Here we discuss VLBI's potential contribution to the MeerKAT extragalactic deep field LSPs. Not all the fields/targets in these surveys are accessible by all stations in existing VLBI Networks (see Fig.~\ref{fig:mutualvis}), however, we proceed in view of only a subset of stations participating for those particular cases and/or in anticipation of potential new stations on the African continent (see \cite{Gaylard2011}).

\subsubsection{MIGHTEE}

Arguably the LSP that stands to benefit most with VLBI coverage (at least from a galaxy evolution perspective) is the MIGHTEE radio continuum and polarimetry survey (Jarvis et al., these proceedings). While MeerKAT's high sensitivity and wide field-of-view will enable deep surveys over large area, these will be limited to $\gtrsim$5~arcsec angular resolution (at 1.4 GHz), which is not sufficient to discern radio jets from star formation in galaxies beyond low redshifts ($z\gtrsim0.2$). This is of key importance for a number of MIGHTEE's science objectives, particularly tracing the star formation rate history of the Universe and understanding the role of mechanical feedback in galaxy evolution (e.g. \cite{McAlpine2015}). One of the most basic, multi-decade old questions is \emph{what fraction of the radio emission from the radio-quiet class (which dominate the radio galaxy total number counts) is associated with AGN activity?} There have been a range of systematic multi-wavelength campaigns to tackle this, with conflicting conclusions (e.g. \cite{Kimball2011, Zakamska2016,White2017}). A direct way to address this problem is through deep, wide-field VLBI observations, an early example of which is shown in \cite{Herrer2016}, who investigate radio-quiet quasars at mas-resolution through a $\sim10~\mu$Jy\,beam$^{-1}$ survey of the COSMOS field and detect compact cores in 3/18 radio-quiet quasars. The power of VLBI observations to discern the primary radio emission component also extends down to the lower luminosity population of star forming galaxies, which dominate the low redshift radio luminosity function ($P_{\rm 1.4GHz} \lesssim 10^{23}$~W\,Hz$^{-1}$, e.g. \cite{Mauch2007}). Higher sensitivity, wider area VLBI observations will increase the statistics and \emph{directly} address these source composition questions.

 Spectral line and continuum VLBI will play an important role in \hi (and OH) emission and absorption components of MIGHTEE, contributing unique morphological insights to the interpretation of gas inflows and outflows in particular (e.g. \cite{Morganti2013}). VLBI polarimetry will not only provide detailed information on sub-kpc magnetic fields and jet physics, but also enable comparison with (and separation from) the larger scale polarisation properties to be probed by MeerKAT (e.g. \cite{Agudo2015}).

The MIGHTEE survey would therefore be significantly enhanced by matched sensitivity VLBI observations across a significant fraction of the selected fields.  To this end, the MIGHTEE-VLBI working group will (a) actively pursue securing open time on global VLBI networks over the course of the next $\sim$5 years over the relevant extragalactic fields, and (b) explore the possibility of capturing the individual MeerKAT antenna data streams simultaneous to the MIGHTEE interferometric observations in order to perform off-line beamforming to retain the wide MeerKAT field-of-view in joint MeerKAT-VLBI observations. The latter is a computationally expensive but scientifically rich route, which we elaborate on in more technical detail in Sec.~\ref{sec:offlinebeam}. Additional survey objectives will include multi-epoch VLBI imaging in order to select sources based on their milli-arcsecond flux variability; and to incorporate e-MERLIN observations where possible to bridge the L-band {\sl uv}-coverage between the MeerKAT and VLBI components, enhancing the interpretation of both. 

\begin{figure}[h]
\centering
\includegraphics[width=0.75\textwidth]{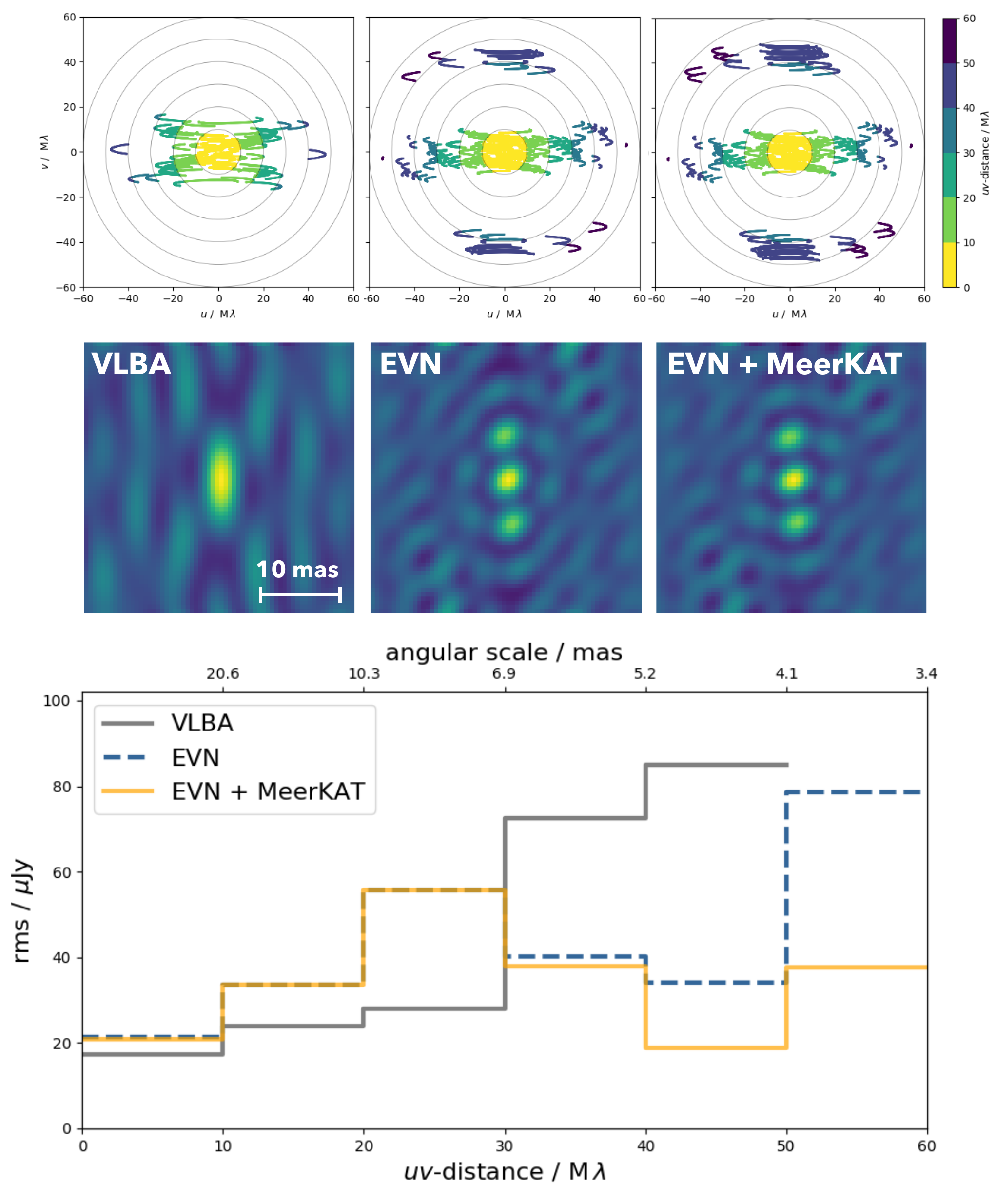}
\caption{Comparison of  the VLBA, EVN and EVN+MeerKAT 18 cm performance towards the XMM-LSS extragalactic field, one of the selected equatorial MIGHTEE fields (Dec. = -4.8~deg). For equivalent comparison, all the above simulations assume 128 MHz of bandwidth per polarization. {\bf Top panel: }Fourier coverage from a 24 hour track. {\bf Middle panel: } Resultant PSFs, each frame with extent 32$\times$32 mas$^2$ and set to the same colour-scale. {\bf Bottom panel: } Sensitivity per \emph{uv}-bin in 10~M$\lambda$ increments, demonstrating the major improvement in $\gtrsim40$~M\,$\lambda$ spatial frequency range with the addition of MeerKAT, despite the lack of full band overlap (which is accounted for). The above plots also indicate VLBI surveys of the XMM-LSS field would be enhanced by the participation of (a subset of) stations in the the upcoming African VLBI Network (AVN). }
\label{fig:uvcov}
\end{figure}

If a wide-field approach is favoured, this naturally places an emphasis on MIGHTEE-VLBI observations in the L-band. However, multi-band VLBI on selected subsamples, the focus of which may be on S-band (and perhaps later X-band) to combine with the corresponding MIGHTEE components, will be an important spectral index and morphological complement to the suite of multi-wavelength, multi-scale MIGHTEE data. A comparison of VLBA, EVN and EVN+MeerKAT performance towards one of the equatorial MIGHTEE fields (XMM-LSS) is shown in Fig.~\ref{fig:uvcov}, demonstrating the dramatic long-baseline sensitivity enhancement made with MeerKAT's inclusion.

\subsubsection{LADUMA}

As described in Baker et al. (these proceedings), LADUMA is a single pointing survey ($\sim 1$~deg$^2$ at 1~GHz) using both UHF and L-bands, centred on Chandra Deep Field South (CDFS). It is a spectral line survey, the primary objective of which is understanding the evolution of neutral hydrogen in galaxies as a function of redshift, stellar mass and environment. Naturally, this will result in a deep (albeit confusion-limited) continuum survey as well and will form part of the MIGHTEE coverage (Jarvis et al., these proceedings), just as the MIGHTEE survey will perform a commensal \hi survey \cite{Maddox2016}. LADUMA's depth will probe down to Milky Way-like galaxies out to redshifts of $z \lesssim 0.5$ and enable a systematic study of neutral hydrogen's presence in starforming and AGN hosts. 

As in the MIGHTEE-HI case (and despite the low declination of the CDFS field), VLBI could play an important role in \hi emission and absorption components of the LADUMA survey, contributing important morphological constraints to the interpretation of gas inflows and outflows (both spectral line and continuum VLBI). This will have strong complementarity with ALMA, which has already shown dramatic examples of this from a molecular gas perspective (e.g. \cite{Combes2013}). By combining these high resolution views of the precursor (\hi) to, and direct fuel (H$_2$) for star formation, a systematic study of the role of radio jets in positive and/or negative feedback would be enabled. The comparable resolution of ALMA (in extended configurations) and VLBI, combined with the morphological information gleaned from \hi emission and absorption, will enable a dramatic increase in the known radio-jet feedback examples such as that presented in \cite{Morganti2013}.

Finally, VLBI may also play an important calibration role with a survey as deep as LADUMA, particularly if the full cumulative visibility database on the field cannot be simultaneously processed. In this case, combination in the image domain will have to be employed. While source structure errors that result from intrinsic time-variability may not affect direct \hi detections, the non-Gaussian nature of the residuals may limit the depth of high-redshift stacking experiments. By monitoring the brightest sources in the MIGHTEE fields (near-)contemporaneously, VLBI could provide the LADUMA calibration pipeline with higher accuracy source models for bright sources which may limit the achievable dynamic range. Uncertainties in source structure can be the limiting factor once certain sensitivity levels are achieved (e.g. JVLA 3C147 field, Smirnov, private communication) and so it may be prudent (and inexpensive) to carry out dedicated higher angular resolution observations of a subset of bright sources. A systematic set of simulations to model of the extent of this effect will be carried out.

\subsubsection{MALS}

The MeerKAT Absorption Lines Survey (MALS) is a $\sim4000$~hour LSP which will search for \hi and OH absorption for redshifts below $z\lesssim 1.8$ (Gupta et al., these proceedings). The objective is to increase the number of known absorbers by more than an order of magnitude ($\sim$600), which will be a formidable sample. Potential followup observations of MALS-discovered absorbers can be separated into two parts: (1) continuum VLBI to discern the fraction of radio flux that is compact and the core/jet geometry; and, where possible, (2) spectral line mode to measure the absorption line at mas-scales. The latter is particularly valuable for ISM physics, constraints on \hi cloud size and opacity. It should provide important examples of positive/negative feedback, again in concert with ALMA, and unveil new insights as to the role of radio jets in \hi to H$_2$ conversion. Since the continuum is already known to be behind the intervening gas, this would provide important constraints on sub-kpc inflows/outflows, not directly attainable by any other means. A MALS-VLBI programme wouldn't necessarily benefit from wide-field capability nor would the contemporaneous MeerKAT-VLBI observations be an efficient use of telescope time, however, a well-chosen MALS sub-sample for VLBI followup should yield a very rich set of results.

\subsubsection{MESMER}
The MeerKAT Search for Molecules in the Epoch of Reionization (MESMER) is an X-band spectral line and continuum survey, primarily aimed at $z = 7-10$ CO~(1-0) detections and characterising the high frequency radio source population \cite{Heywood2011}. While the MeerKAT X-band receivers remain unfunded and high redshift CO (1-0) detection prospects have decreased since the original 2010 proposal (due to new observational constraints, theoretical progress, and decreased instantaneous bandwidth), MESMER remains, at the very least a unique deep continuum survey which is expected to have excellent legacy value into the SKA era. Tier~3 will cover 1~deg$^2$ (likely over CDFS) down to $\sim$1~$\mu$Jy\,beam$^{-1}$, and will be a definitive survey in terms of understanding the nature of the faint source radio population at high frequency. Similarly to the radio-quiet quasars and star forming galaxies, a major question is what fraction of this radio emission is in the form of compact, high brightness temperature AGN emission? Source counts at higher frequencies are significantly under-predicted (based on low-frequency counts) below flux densities of $\sim 1$~mJy\,beam$^{-1}$ (e.g. \cite{AMI2011}). Recent results by \cite{Whittam2013} reveal a population of flat-spectrum sub-mJy sources at 15.7~GHz. Their existence is not predicted by semi-analytic source population models (e.g. \cite{Wilman2008}, and so they could be young objects; frustrated FR-I's with little to no jet/lobe emission; or a new, unaccounted for class of objects. All of these cases would be effectively investigated with extensive VLBI coverage over the MESMER footprint to systematically determine the fraction that host compact radio emission at GHz frequencies, and if any mas-scale morphological insights may be gleaned. Southern African VLBI stations on $500-1000$~km baselines would be highly advantageous in this regard, as their lower brightness temperature sensitivity would enhance high fidelity mapping of sub-arcsecond jet features.

\section{Technical development plan}

Efforts to enable MeerKAT VLBI capability are actively underway through the MeerKAT-VLBI working group.  Given the significant system overlap between KAT-7 and MeerKAT, a dedicated VLBI observing mode is being prototyped on the former. A more detailed description of the technical plan, as well as initial results with the EVN, will be outlined in an upcoming MeerKAT-VLBI White Paper. Here we discuss technical capabilities and possibilities relevant to wide-field VLBI surveys in particular.

\subsection{Commensal VLBI observations}

MeerKAT's array configuration has maximum baselines of 8~km. With its pinched core array configuration (50 percent of the collecting area within 1 km), the angular resolution ranges between 10-15~arcsec for Briggs-weighted images with robust parameters from 0.5-2. Achieving better angular resolution than this would effectively require down-weighting the array core to such an extent that there would likely be little difference if some antennas in the core were simply not included in the observation. Mauch et al. (in prep.) have considered this with final results pending, however, clearly there is an opportunity to optimise the scientific output of MeerKAT, by ensuring the full use of the array core is scientifically and technically justified for open time MeerKAT proposals. In those cases where the full core is not required, a subarray could be used to participate in VLBI observations. Intelligent dynamic scheduling could synchronise such cases with the EVN/LBA sessions such that a subset of MeerKAT core antennas could be available, should an approved MeerKAT-VLBI (or other) observing programme be possible to carry out. As indicated previously, MeerKAT will be able to operate up to four sub-arrays, each of which could simultaneously operate in interferometric and tied-array beam mode, with up to four independent beams in the latter. This is obviously a flexible system, which could lead to highly efficient commensal VLBI operations in both open and LSP time. This is not to suggest that it should be a standard operating mode, but rather just to present an opportunity where the scientific output of MeerKAT can be further increased.

\subsection{Offline beamforming}\label{sec:offlinebeam}

In tied-array beam mode, MeerKAT will form up to four independent beams in real time. The limitation to four independent beams is due to the computational (and hence financial) expense. However, one could in principle record the individual antenna data streams and beamform offline -- an idea that was put forward during the MeerKAT Science Workshop by Russ Taylor. Without the real-time processing constraints, one could store all the data for a $\sim$1000-2000 hour survey like MIGHTEE and beamform offline to cover a larger fraction of the combined VLBI array's field-of-view. \emph{This would have the potential to increase the processed field-of-view by over 2-3 orders of magnitude all for the cost of additional computing, rather than additional telescopes and/or VLBI time}. Moreover, assuming participating VLBI stations would approve the project, it would cost zero extra MeerKAT time since it would be carried out during already scheduled MIGHTEE survey observations. If African stations with similar wide-band coverage to MeerKAT were to be built, this would provide simultaneous \hi absorption opportunities for a subset of the participating VLBI stations, as well as improved \emph{uv}-coverage due to the large fractional bandwidth. 

While offline beam forming would not necessarily enjoy the low scaling factor of multi-source phase centre technique, the correlator required would be sub-MeerKAT and could be carried out over a timescale many times that of real time, thereby lower the processing requirements. The technical feasibility and cost implications are currently being explored by the MeerKAT-VLBI Working Group and will be reported on in an upcoming White Paper.

\subsection{Wide-field imaging and source finding}

As discussed in Sec.~\ref{sec:intro}, the past decade has seen significant advances in wide-field cm-wave VLBI surveys through increased instantaneous bandwidth, the multi-phase centre correlation technique, and multi-view calibration (e.g. \cite{Rioja2009}). This increased wide-field VLBI capability is timely given the survey-driven focus of next-generation radio interferometers, however, poses significant post-processing challenges which can curtail the potential scientific yield. Fortunately, there are many developments in the km-scale connected-element interferometer sphere that are directly applicable to wide-field VLBI challenges. Deane et al. (in prep.) demonstrate the use of a wide-field imager designed for low frequency, compact arrays such as the MWA (\cite{Offringa2014}), to make 205 x 64k x 64k, ($\sim$1 Terapixel combined) VLBI images in just 1-2 weeks of processing time on a standard 32 core compute server without any optimisation -- far exceeding the performance of imagers in more standard software packages. This is part of a quasi-uniform sensitivity VLBA survey of the GOODS-North field and is used as a comparison to the more standard technique of placing phase centres on sources already detected at $\sim$arcsec angular resolution. Deane et al. also demonstrate the use of a number of other arcsec-scale algorithms which accelerate and introduce more flexibility to standard VLBI post-processing. There are many other examples of more efficient, easy-to-use/modify software that is now being employed in a VLBI context, from fringe-fitting to source finding. The key point is that many of the post-processing developments that have been driven by large data volume connected-element surveys will directly benefit wide-field VLBI efforts.

\section{Summary}

There are a wide range of high profile science cases that would be greatly enhanced/enabled by the addition of an effective 110 metre dish on 7,000+ km baselines to global VLBI networks. Many of these would only require traditional, targeted VLBI observations, so the small field-of-view afforded by a phased up MeerKAT array would not be detrimental. Comparable sensitivity VLBI coverage over some of the accessible fields would significantly enhance the MeerKAT LSP scientific output (e.g. MIGHTEE, LADUMA, MALS, MESMER). This could be in the form of targeted follow-up surveys of carefully selected subsamples, however, the telescope efficiency and discovery space would be dramatically increased if offline beamforming were to be possible for certain MeerKAT LSPs. This would be an important technical and scientific step forward towards full SKA (and SKA-VLBI) operations.

\section*{Acknowledgements}
\noindent The author thanks Griffin Foster, Justin Jonas, Mike Garrett, Fernando Camilo, Oleg Smirnov, Lindsay Magnus, Sean Passmoor, Simon Ratcliffe and the MeerKAT-VLBI Working Group for fruitful discussions.

\end{document}